\documentclass[prl,twocolumn,superscriptaddress]{revtex4}

\usepackage{amsmath} 
\usepackage{amssymb} 
\usepackage{amsfonts}

\usepackage{graphicx} 
\usepackage{color}

\begin{document}

\bibliographystyle{apsrev4-1} 

\title{Curvature Dependence of Hydrophobic Hydration Dynamics} 

\author{R. Gregor Wei\ss}
\affiliation{Department of Physics, Humboldt Universit{\"a}t zu Berlin, Newtonstr.~15, D-12489 Berlin, Germany}
\affiliation{Soft Matter and Functional Materials, Helmholtz-Center Berlin, Hahn-Meitner Platz 1, D-14109 Berlin, Germany}
\author{Matthias Heyden}
\affiliation{Max-Planck-Insitut f{\"u}r Kohlenforschung, Kaiser-Wilhelm-Platz 1, D-45470 M{\"u}lheim an der Ruhr, Germany}
\author{Joachim Dzubiella}
\thanks{To whom correspondence should be addressed. E-mail: joachim.dzubiella@helmholtz-berlin.de}
\affiliation{Department of Physics, Humboldt Universit{\"a}t zu Berlin, Newtonstr.~15, D-12489 Berlin, Germany}
\affiliation{Soft Matter and Functional Materials, Helmholtz-Center Berlin, Hahn-Meitner Platz 1, D-14109 Berlin, Germany}

\begin{abstract}
We investigate the curvature-dependence of water dynamics in the vicinity of hydrophobic spherical solutes 
using molecular dynamics simulations.  For both, the lateral and perpendicular diffusivity as well as for H-bond kinetics  of water in the first hydration shell, we find a non-monotonic solute-size dependence, exhibiting extrema 
close to the well-known structural crossover length scale for hydrophobic hydration.  Additionally, we find an 
apparently anomalous diffusion for water moving parallel to the surface of small solutes, which, however, can be explained
by topology effects. The intimate connection between solute curvature, water structure
and dynamics has implications for our understanding of hydration dynamics at heterogeneous 
biomolecular surfaces.

\end{abstract}

\maketitle

%{\it Introduction}-- 
One of the greatest advances in our understanding of the hydrophobic effect is the recognition that 
the hydration structure and thermodynamics of apolar solutes is qualitatively length scale dependent~\cite{Chandler, Lee&Rossky, Honig, Lum&Weeks, Huang&Chandler2001, Huang&Chandler2002, Rajamani&Garde2005, Ashbaugh&Pratt, Pratt, Djikaev&Ruckenstein, Davis&Ben-Amotz}. 
The microscopic reason is that water structures very 
differently at small (convex) solutes, where the bulk H-bond network is only moderately
deformed, as compared to large solutes, which significantly distort the tetrahedral bulk structure. 
The structural crossover happens at sub-nanometer length scales and has important implications for the interpretation of the structure and thermodynamics of 
hydrophobically-driven assembly processes~\cite{Chandler, Berne&Weeks}, such as protein 
folding and association~\cite{Huang&Chandler2000, Athawale&Garde, Li&Walker}. 

The dynamics of the hydration layer that surrounds molecular self-assemblies and proteins in solution 
has attracted plentiful interest in the last decade~\cite{bagchi}. Solute fluctuations and hydration dynamics are understood to be highly coupled with important
consequences to biological function, such as enzyme catalysis and molecular recognition in 
binding~\cite{bagchi, Niehues&Havenith, Kwon&Zewail, Zhang:PNAS, Pizzitutti, Fogarty, Setny:PNAS,  Mittal&Hummer, Geissler, Amish, Jamadagni}. 
Despite the obvious importance of the solute chemical composition, 
apparently the intrinsic topological and geometric features 
play an important role as well~\cite{Zhang:PNAS, Cheng&Rossky, Pizzitutti, Fogarty}, possibly even leading  
to anomalous diffusion behavior~\cite{Pizzitutti}. Therefore, and due to the established fact that water considerably 
restructures at radii of curvature close to the sub-nanometer scale,  
a natural question to ask is, {\it how does the water structural crossover affect the dynamics of the hydration 
layers in the solute vicinity?} 

One of the first simulation studies of curvature effects on hydration dynamics was performed by Chau {\it et al.} for three solute 
radii between 0.35 and 0.8~nm~\cite{Chau&Smith}. They found 
a slowdown of the diffusion of water in the first hydration shell relative to the bulk with 
an apparent minimum for  the intermediate solute size. This interesting finding was not commented on, probably due to the little amount of data
and statistical uncertainty of the results. Further, a slowdown of  water reorientational dynamics was found  compared to bulk, an 
effect that decreased with solute size~\cite{Chau&Smith}.  The reorientational slowing-down has been recently explained  by excluded-volume  effects on the H-bond exchange dynamics~\cite{Laage&Hynes2008, Galamba}. In disagreement with the simulations with Chau {\it et al.}, however,  
the excluded-volume concept predicts a monotonic increase of the reorientation times~\cite{Laage&Hynes2008}, 
an apparent controversy which has not been addressed in literature, yet. 

In this letter, we report on systematic molecular dynamics (MD) simulations of water around hydrophobic spherical 
model solutes of varying radius between 0.3 and 2.1~nm  and investigate intrinsic curvature effects on water hydration dynamics.  We find that the perpendicular and lateral diffusivity of hydration water exhibit a non-monotonic
curvature dependence, with temperature-dependent minima located close to the structural 
crossover length scale~\cite{Rajamani&Garde2005, Ashbaugh}. Both, curvature and temperature dependence, are
strikingly similar to that of entropy during hydrophobic solvation~\cite{Ashbaugh} and thus allows strong conjecture that
the hydrophobic effect extends beyond thermodynamic into dynamic anomalies.
Furthermore, we find that the intriguing curvature dependence of the diffusivity is found to be related to a non-monotonic 
dependence of H-bond life times on curvature, exhibiting a maxima at the crossover length scale. This finding 
reconciles previous, apparently contradicting results from simulations ~\cite{Chau&Smith} and 
theory~\cite{Laage&Hynes2008, Galamba} 
on reorientation times, as we find both predicted behaviors but in  different length scale regimes.  
The intimate connection between solute curvature and water structure and dynamics 
should be of fundamental importance for biomolecular function mediated by heterogeneous biomolecular surfaces~\cite{bagchi, Niehues&Havenith, Zhang:PNAS, Cheng&Rossky, Pizzitutti, Fogarty}.

%{\it Simulation Methods}--
Our MD simulations are performed each containing a single and fixed hydrophobic model solute solvated in SPC/E water~\cite{SPCE} using the Gromacs 
simulation package~\cite{gromacs}.
The solute-water interaction is mediated by a shifted 
Lennard-Jones potential $U_{LJ}(r') = 4\epsilon[(\sigma/r')^{12}-(\sigma/r')^{6}]$, whereas $r' = r - r_0$  describes a
coordinate shift by $r_{0}$. With such a shifted potential, it is guaranteed that by changing the solute size {\it only} 
intrinsic curvature effects are probed, not, e.g., those of an additionally varying dispersion attraction.
 \begin{figure}[t]
	\includegraphics[scale=1]{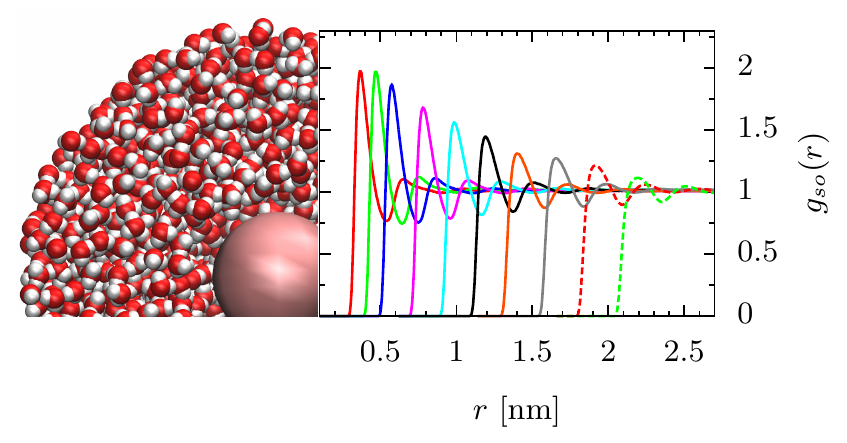}
\caption{Left: MD simulation snapshot of a hydrophobic sphere with shift radius $r_0 = 8\,\text{\AA}$ in explicit water. Right:
Radial distribution function (RDF) $g_{so}(r)$ of water around differently sized hydrophobic model solutes.}
\label{snapshot_rdf}
\end{figure}
Values of $\sigma=3.768$~\AA~ and $\epsilon=1.197$~kJ/mol are chosen from a model for methane 
to mimic  a reasonable dispersion attraction with a well-defined first solvation shell~\cite{Huang&Chandler2002}. 
We choose shift radii $r_{0} = 0\,\text{\AA}, 1\,\text{\AA}, 2 \,\text{\AA}, 4\,\text{\AA}, 6\,\text{\AA}, 
8\,\text{\AA}, 10\,\text{\AA}, 12.5\,\text{\AA}$, $15\,\text{\AA}$ and $17.5\,\text{\AA}$ in ten separate simulations. 
As a limiting case of a hydrophobic surface with zero curvature, the box delimiting walls in the $x-y$-plane in a
pseudo-2D simulation are chosen to interact with
water by a 12-6 potential in $z$-direction $U_{12-6}(z) = 4\epsilon[(\sigma/z)^{12}-(\sigma/z)^{6}]$.
After $100\,\mathrm{ps}$ of Gibbs ensemble ($NPT$) equilibration,  canonical ($NVT$) production runs are 
performed up to $200\,\mathrm{ns}$ at a temperature of $T = 300\,\mathrm{K}$ and with $N=6000$ to 12000 water molecules.  
Further details on the simulation technique can be found in the Supporting Information (SI).
A simulation snapshot and water oxygen density profiles around all the solutes are presented
in Fig.~1.  

%{\it Diffusion Dynamics in the Hydration Shell}
We first characterize the water diffusion parallel to the solute surface. For this, we calculate the lateral mean square
displacement (MSD)
of the arc length a water oxygen has traveled in time $t$, via
\begin{equation}
 \langle S(t)^2 \rangle = R_{\rm avg}^2\cdot\langle [\theta(t' + t) - \theta(t')]^2 \rangle_{R_{1}} \,\mathrm{.}
 \label{eq:MSD}
\end{equation}
Here $R_{\rm avg} = \int_{0}^{R_{1}} \mathrm{d}r\, r\cdot g_{so}(r)$ denotes the average distance of the water molecules inside the first hydration layer, which is delimited by the location $R_{1}$ of the first minimum in the RDF (cf.~Fig.1), to the solute center. The time average $\langle...\rangle_{R_1}$ over starting times $t'$ is taken  only in the first hydration shell. The variable $\theta(t)$ is the
azimuthal angle of the water oxygen-solute distance vector at time $t$.  For water molecules diffusing at a wall,  Eq.\eqref{eq:MSD} 
reduces to the usual two dimensional MSD $\langle [\Delta x(t) + \Delta y(t)]^2
\rangle_{R_{1}}$ in the $x-y$-plane. 

\begin{figure}[t]
	\includegraphics[scale=1]{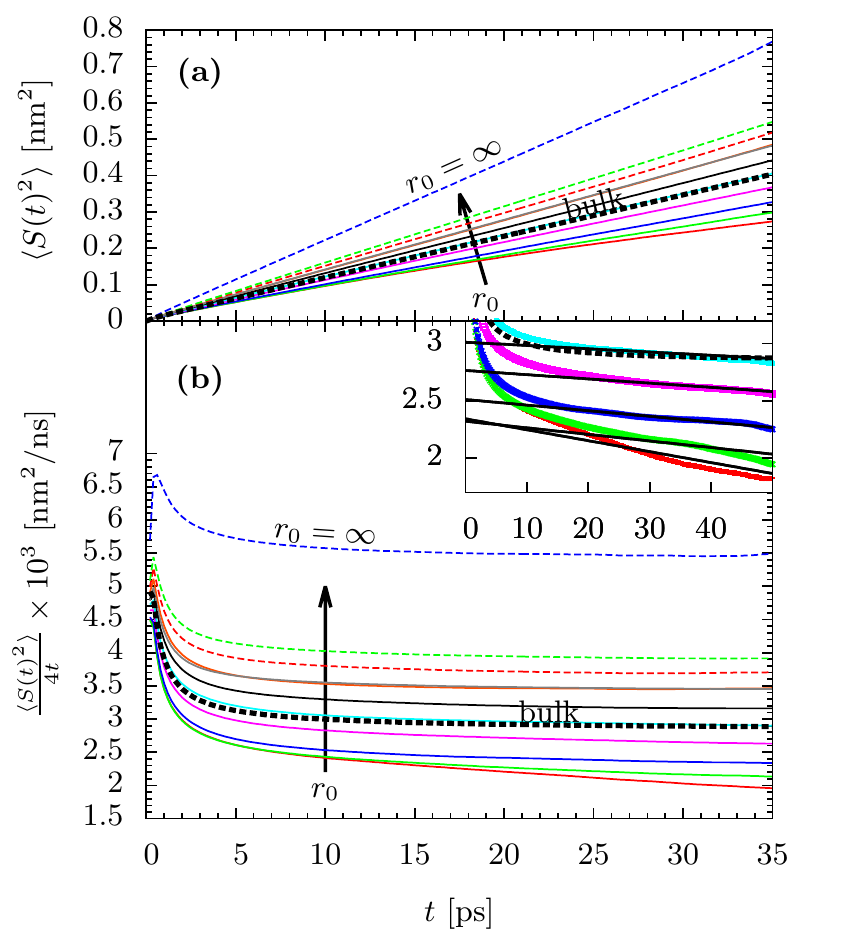}
\caption{ (a) MSD of water molecules parallel to differently curved hydrophobic surfaces and MSD of water in bulk (thick black
dashed line). (b) The data from the top panel (a) divided by $4t$. The \textit{Inset} shows the curves for water at the five smallest
solutes along with the corresponding fits from Eq.~(\ref{eq:MOMENT}.)}
\label{fig:MSD}
\end{figure}

The lateral MSDs for water around all solutes are shown in Fig.~\ref{fig:MSD}(a) along
with those of water at the planar hydrophobic surface and in bulk.  Comparison to the bulk MSD indicates a crossover from below to above bulk water
self-diffusion with increasing slopes with growing solute size. Additionally, the hydrating water of the smallest solutes exhibits non-linear, anomalous behavior 
in $t$. This can be better recognized by the negative slope for small solutes in Fig.~\ref{fig:MSD}(b) where we plot the MSD divided by time, 
$\langle S(t) \rangle / 4t$. Linearity of the MSDs  is restored for larger solutes, converging 
towards the limiting MSD $r_0 \rightarrow \infty$ of the planar hydrophobic surface. 

The apparently anomalous diffusion behavior for small solutes can be explained by intrinsic curvature effects, which
modify the standard 2D diffusion law~\cite{Gosh&Sinha}. Here, the probability distribution function (PDF) of diffusion
on spherical surfaces reads
\begin{equation}
 P(\theta,\tau) = \frac{N(\tau)}{\tau}\sqrt{\theta\mathrm{sin}(\theta)}\cdot\mathrm{e}^{-\frac{\theta^2}{2\tau}} \, 
 \label{eq:PDF}
\end{equation}
where $\tau = 2D_{||}t/R_{avg}^2$, and $N(\tau)$ is a normalization constant. By expanding Eq.~\eqref{eq:PDF} up to second order for
small $\theta$, we find that the second moment of the PDF 
can be written for small solutes as
\begin{equation}
 \langle \theta^2 \rangle R_{avg}^2 \approx 4D_{||}t-\frac{4}{3}\frac{D_{||}^2 t^2}{R_{avg}^2} \,\mathrm{.}
 \label{eq:MOMENT}
\end{equation}
On surfaces with high curvature ($R_{avg}^2 \ll D_{||}t$) the second term on the r.h.s. of Eq.~\eqref{eq:MOMENT} slows down the MSD as found in our simulations. In the limit $R_{avg} \rightarrow \infty$,  this term vanishes and the MSD on planar surfaces,  $\propto 4D_{||}t$,  is restored, 
consistent with the vanishing non-linearity as observed in Fig.~\ref{fig:MSD}. 

\begin{figure}[t]
\vspace{-0.7cm}
\centering
	\includegraphics[scale=1]{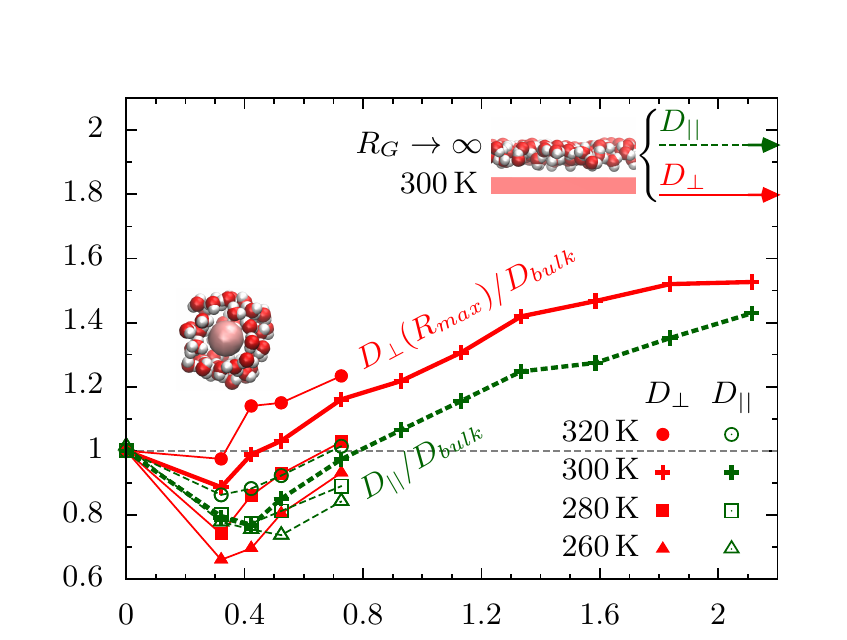}
\caption{Parallel and perpendicular diffusivities $D_{||}$ (green dashed lines) and $D_\perp$ (red solid lines) scaled by bulk water diffusion 
$D_{bulk}$ against solute size $R_{G}$ for $T=300$~K. For the four smallest cavities the temperature effect on $D_{\perp}$ and $D_{||}$ is shown for
$T=260\,$K, $280\,$K and $320\,$K.
The grey dashed horizontal line represents bulk water diffusivity.
The values corresponding to diffusivity at the limiting case of zero curvature ($R_{G}\rightarrow\infty$, only $T=300\,$K) are drawn as horizontal arrows with respective color coding.}
\label{fig:DiffConst}
\end{figure}

By fitting Eq.~\eqref{eq:MOMENT} to our MSDs, we obtain the lateral diffusion constant $D_{||}$, 
see the inset in Fig.~\ref{fig:MSD}(b) for the fits.  The results are plotted versus the Gibbs radius of the solutes in Fig.~\ref{fig:DiffConst} at $T=300$~K normalized by bulk diffusion constant $D_{bulk}$ of water.   For small solutes, we find that the  parallel diffusion $D_{||}$  slows down with increasing solute sizes which leads to parallel mobilities smaller than in
bulk. This  trend changes  at a minimum at a length scale of $R_{G}^*\approx 0.45\,\mathrm{nm}$ beyond which
the lateral diffusivity monotonically rises with decreasing surface curvature to become bulk-like at about $R_{G}^*\approx 0.75\,\mathrm{nm}$  and finally saturates to a water mobility higher than in bulk, as well known 
for  smooth and planar hydrophobic surfaces~\cite{Kumar}.  The significant length scales observed here match well the 
structural  crossover length scale of $\simeq 0.5$~nm found for SPC/E water estimated from solvation 
free energies of spherical model solutes~\cite{Rajamani&Garde2005}. 

\begin{figure}[t]
\centering
	\includegraphics[scale=1]{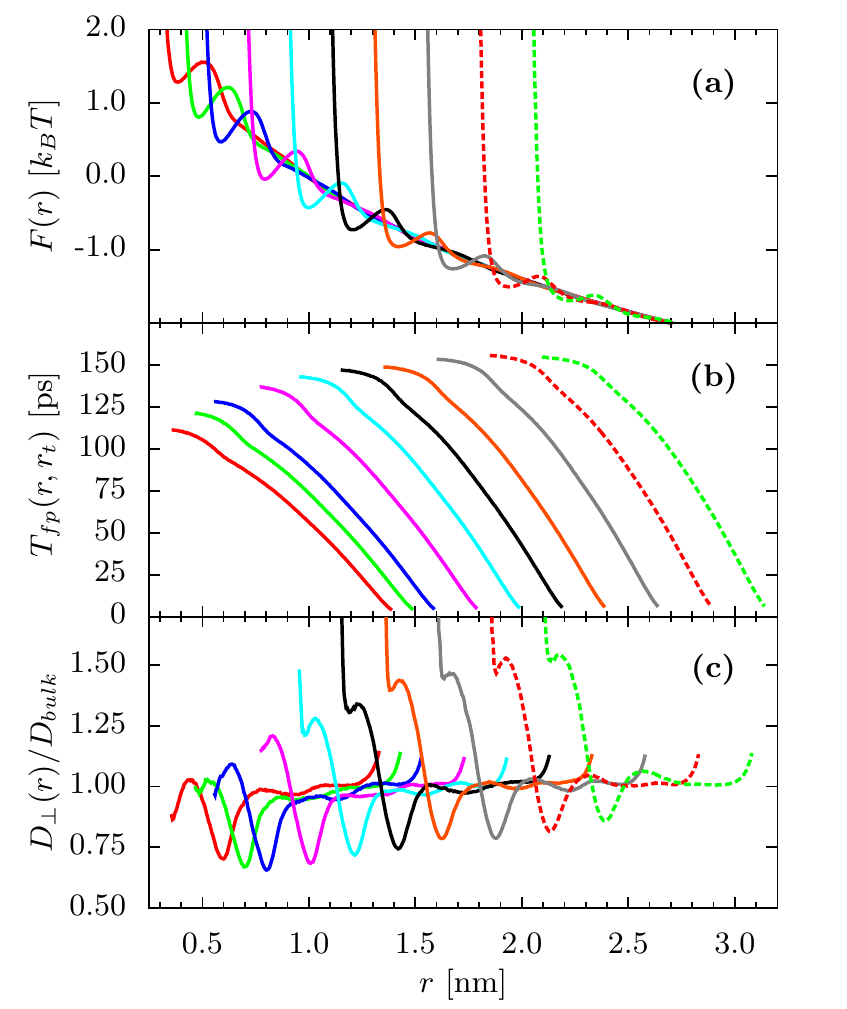}
\caption{(a) Free energy profile $F(r)$ for water around differently sized hydrophobic solutes obtained by the Boltzmann
inversion of the water RDFs in Fig~1. (b) MFPT curves $T_{fp}(r,r_t)$ for water
molecules to reach the target distance at $r_t = 14\,\text{\AA}$ given they started at a distance $r<r_t$. (c)
Perpendicular diffusivity profiles $D_{\perp} (r)$ obtained from Eq. \eqref{eq:perpDiff}.}
\label{fig:MFPTmethod}
\end{figure}

To calculate the water diffusivity {\it perpendicular} to the hydrophobic surfaces, we employ the mean first-passage time
(MFPT) analysis introduced by Hinczewski {\it et al.} \cite{Hinczewski&Netz,Sedlmeier&Netz}. 
The MFPT $T_{fp}(r,r_{t})$ describes the mean time required by a molecule to travel from distance $r$ to the solute to a target distance $r_{t} > r$ and has an exact solution in a Smoluchowski description in terms of the free energy
$F(r)$ and diffusivity profile $D_{\perp}(r)$ \cite{Weiss}. Its inversion leads to an
expression for the perpendicular diffusivity profile~\cite{Hinczewski&Netz,Sedlmeier&Netz}
\begin{equation}
 D_{\perp}(r) = -\frac{\mathrm{e}^{\beta F(r)}}{\partial T_{fp}(r,r_{t})/\partial r} \int_{r_{min}}^{r}\,\mathrm{d}r'\,\mathrm{e}^{-\beta F(r)}
 \label{eq:perpDiff}
\end{equation}
with $\beta^{-1} = k_{B}T$ and $r_{min}$ being a reflective boundary close to the solute where $F(r_{min}) = 10\,k_{B}T$.
From our MD simulations, we extract MFPTs $T(r,r_{t}=r_{0}+14\,\text{\AA})$ as shown in the
middle panel (b) of Fig.~\ref{fig:MFPTmethod} along with free energy profiles $F(r)$ in the upper panel (a). 
The latter result from a simple Boltzmann inversion of the solute-water RDFs plus the 
entropic contribution $F(r) = -k_{B}T\,\mathrm{ln}\,g_{so}(r) -2k_{B}T\,\mathrm{ln\,}r$~\cite{Sedlmeier&Netz}.  The bottom panel (c) plots the resulting diffusivity profiles
$D_{\perp}(r)/D_{bulk}$ normalized by bulk diffusivity for all solute sizes. The general shape of the profiles describes a maximum 
inside the first solvation shell followed by
a minimum in diffusivity which is reached at the outside margin of the solvation layer, whereafter diffusivity converges towards
bulk diffusion in an oscillatory fashion. (The final rise of the profiles for radii close to the target distance is an
artifact of non-Markovian contributions~\cite{Hinczewski&Netz,Sedlmeier&Netz}, see also the SI for technical details.) 
In our calculated profiles,  an interesting non-monotonic curvature dependence of the perpendicular water mobility near hydrophobic
surfaces is visible in the extrema of the profiles.  With growing solute size,  the maximum of diffusivity in the first hydration shell continues to increase which
leads to high perpendicular diffusivity in vicinity of weakly curved surfaces, whereas the minimum becomes less pronounced and
almost vanishes in the profile at the planar limit (see SI). 

We now define the perpendicular diffusion coefficient $D_{\perp}(R_{max})$ in the first hydration shell by the value of the diffusivity curve at $R_{\rm max}$, the 
position of the first peak of the RDF. The scaled value $D_{\perp}(R_{max})/D_{bulk}$ is presented next to the scaled values of $D_{||}$ in 
Fig.~\ref{fig:DiffConst} versus solute size $R_G$ for $T=300$~K. Like $D_{||}$, the perpendicular diffusivity also shows the crossover from
below-to-above bulk mobility, but at a smaller distance of about 0.4~nm, indicating a preceding non-monotonic solute size dependence for small solutes as it is
interpolated towards the limiting (bulk) case of a solute with vanishing radius.  
Since the magnitude of the perpendicular diffusivity slightly depends on  the definition 
of what radius defines the first hydration shell, or if over all water molecules in this shell should be averaged, we compare those
different definitions in the SI.  We find similar values, in particular, all tested definitions rigorously reproduce the crossover from below-to-above bulk mobility behavior of water versus solute size close to the structural crossover length scale.

Additionally Fig.~\ref{fig:DiffConst} plots diffusivity changes upon three different temperatures $T=260\,$K,
$280\,$K and $320\,$K at cavities at which the dynamic anomaly occurs. Decreasing temperature shifts the curves and the dynamic
anomaly towards larger radii of curvature. Hence, the observed temperature dependence of diffusive dynamics obeys the same trend
as the thermodynamic and structural crossover length scale~\cite{Ashbaugh}.
Taking into account entropy scaling laws for diffusion~\cite{Rosenfeld,Dzugutov}, which scale
exponentially with excess entropy, suggests entropy to be a constitutive measure for the dynamic anomaly, thus firmly corroborating our conclusions.

Ultimately, water hydrogen bond (HB) kinetics and mobility are intimately linked because translational diffusion is accompanied
by processes breaking, forming and re-forming HBs. The cooperation of elemental dynamic processes in water is successfully described with
reaction-diffusion models \cite{Luzar&Chandler,Luzar,Markovitch&Agmon} and was probed by
simulations~\cite{Luzar&Chandler,Luzar,Markovitch&Agmon,Pereyra&Carignano}.  
Here, we probe HB lifetimes 
$\tau$ which are estimated by the correlation
function $c(t) = \langle h(t) h(0) \rangle / \langle h^2 \rangle$ 
of the HB operator $h(t)$, which is $1$ for a specific pair of water molecules while they are bonded, and $0$
otherwise~\cite{Luzar&Chandler,Luzar,Markovitch&Agmon}. The negative derivative $k(t) = - \mathrm{d}c(t) / \mathrm{d}t $ is the
reactive flux hydrogen bond correlation function. We estimate 'intermittent' HB lifetimes by the zero
frequency part \cite{Luzar}, the integral of $c(t)$,
\begin{equation}
\tau = - \int_{0}^{\infty} t \cdot k(t)\, \mathrm{d}t = \int_{0}^{\infty} c(t)\,
\mathrm{d}t\,\, .  
\end{equation}
It is significantly influenced by diffusion leading to the separation of
initially bonded water molecule pairs after HB breaking. Without this separation, breaking a HB between two water molecules is reversible and the
kinetics associated with this process are much faster than the irreversible rearrangements of the water hydrogen bond network analyzed here
~\cite{Luzar}.
The correlation functions and details on the H-bond definition and numerical evaluation can be found in the SI.

\begin{figure}%[t]
\centering
	\includegraphics[scale=1]{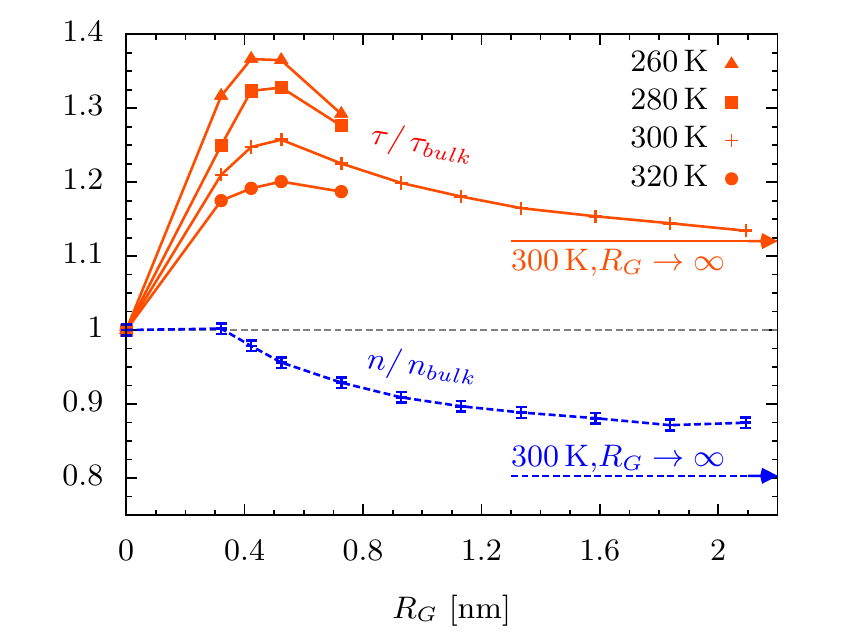}
\caption{Mean hydrogen bond (HB) lifetime of first-solvation shell water $\tau(R_G)$ versus solute size scaled by the bulk value $\tau_{bulk}$ (orange symbols).  
Average number of HBs per water molecule $n(R_G)$ scaled by its bulk value $n_{bulk}$ (blue symbols). 
The values corresponding to the respective measures at the limiting case of zero curvature ($R_G\rightarrow\infty$, only $T=300\,$K) are drawn as horizontal arrows with respective color coding. The errorbars for $n/n_{bulk}$ were estimated from block averages as
described in the SI.}
\label{fig:tau}
\end{figure}

The results for $\tau(R_G)$ scaled by their bulk value
$\tau_{bulk} = 3.4\,\mathrm{ps}$ are plotted in Fig.~\ref{fig:tau}. They exhibit a 
non-monotonic size dependence with a maximum retardation of about 25$\%$ versus bulk at roughly
$R_G\simeq 0.45$~nm, close to the structural crossover length scale.   After the maximum, the
retardation decreases for solute sizes $R_{G} > 0.5\,\mathrm{nm}$ down to a remaining $\approx 5\%$ near the planar surface.
Concurrently,  the average number of HBs per water molecule $n/n_{bulk}$, also shown in  Fig.~\ref{fig:tau}, 
decreases monotonically. Hence, the non-monotonic translational mobility behavior of water 
is solely reflected in the kinetic behavior of the HBs.  The observed behavior 
reconciles previous, apparently contradicting results from simulations ~\cite{Chau&Smith} and theory~\cite{Laage&Hynes2008, Galamba} 
on water reorientation times, as we find both predicted trends (increasing and decreasing life times), 
but in different solute size regimes, roughly separated by the important structural crossover scale. 

Fig.~\ref{fig:tau} also plots the $T$-dependence on the HB correlation times where 
a significant slowing down with decreasing temperature can be observed with not much change of the
position of the maximum on the solute-size axis. We note that the latter behavior cannot be related 
one-to-one  to our observed diffusivity behavior in (Fig~\ref{fig:DiffConst}) due to the multiple timescales entering $\tau$ (\cite{Luzar} and SI).

In summary, we have established a firm link between structure and dynamics of hydration water around 
hydrophobic solutes with a novel dynamic anomaly happening at the well-know crossover length scale. 
Due to the fundamental importance of surface water to biomolecular processes and function~\cite{bagchi, Jamadagni}, 
in particular at topologically heterogeneous protein surfaces~\cite{Zhang:PNAS, Cheng&Rossky, Pizzitutti, Fogarty}, our findings
imply that nature has the means to employ local surface topology to mediate biological function. Hence, our results will help in
the interpretation of experimentally found dynamic heterogeneities on biomolecular surfaces~\cite{bagchi, Niehues&Havenith,
Kwon&Zewail, Zhang:PNAS, Pizzitutti, Fogarty}.  Locally slowed down water, for instance, could thus 
fine-tune the folding kinetics of hydrophobic polymers  and peptides~\cite{bagchi} or may mediate the appropriate time scales  for the association of ligands to catalytically active sites or binding pockets~\cite{bagchi, Grossmann, Setny:PNAS}.

\acknowledgments

The authors thank Roland R. Netz for helpful discussions and the Deutsche Forschungsgemeinschaft (DFG) 
for financial support. M. Heyden is grateful for support by the Cluster of Excellence RESOLV (EXC 1069) funded by the Deutsche
Forschungsgemeinschaft. 

 \end{document}

\bibliographystyle{apsrev4-1} 

\title{Supporting Information for: ''Curvature Dependence of Hydrophobic Hydration Dynamics''} 

\author{R. Gregor Wei\ss}
\affiliation{Department of Physics, Humboldt Universit{\"a}t zu Berlin, Newtonstr. 15, D-12489 Berlin, Germany, Germany}
\affiliation{Soft Matter and Functional Materials, Helmholtz-Center Berlin, Hahn-Meitner Platz 1, D-14109 Berlin, Germany} 
\author{Matthias Heyden}
\affiliation{Max-Planck-Insitut f{\"u}r Kohlenforschung, Kaiser-Wilhelm-Platz 1, D-45470 M{\"u}lheim an der Ruhr, Germany}
\author{Joachim Dzubiella}
\thanks{To whom correspondence should be addressed. E-mail: joachim.dzubiella@helmholtz-berlin.de}
\affiliation{Department of Physics, Humboldt Universit{\"a}t zu Berlin, Newtonstr. 15, D-12489 Berlin, Germany, Germany}
\affiliation{Soft Matter and Functional Materials, Helmholtz-Center Berlin, Hahn-Meitner Platz 1, D-14109 Berlin, Germany}

\maketitle

\subsection{Molecular Dynamics Simulations}

The systems comprise SPC/E water as solvent and a single hydrophobic sphere fixed in the middle of the simulation box. The
interaction between SPC/E and the model solute is mediated by a shifted Lennard Jones potential $U_{LJ}(r') =
4\epsilon[(\sigma/r')^{12}-(\sigma/r')^{6}]$ whereas $r' = r - r_0$ increases the solute size. The parameters $\epsilon =
1.197\,\text{kJ/mol}$ and $\sigma = 3.768\,\text{\AA}$ are those of a united-atom methane molecule as in
Ref.\cite{Huang&Chandler}. Ten seperate simulations are
performed with shift radii $r_0 = 0\,\text{\AA}$, $1\,\text{\AA}$, $2 \,\text{\AA}$, $4\,\text{\AA}$, $6\,\text{\AA}$,
$8\,\text{\AA}$, $10\,\text{\AA}$, $12.5\,\text{\AA}$, $15\,\text{\AA}$ and $17.5\,\text{\AA}$. For shift radii $r_0 \leq
6\,\text{\AA}$ the simulation box containes 6000 SPC/E water molecules and for simulations with $r_0 > 
6\,\text{\AA}$ the number of solute molecules is doubled to 12000 to avoid finite size effects. 

To model the corresponding hydrophobic surface without curvature the walls in the x-y-plane of a simulation box of size
$5.67\times5.67\times6.54\,\mathrm{nm}^3$ are set to interact with water with a 12-6 potential in the z-direction
$U_{12-6}(z) = 4\epsilon[(\sigma/z)^{12}-(\sigma/z)^{6}]$ with same interaction parameters $\epsilon = 1.197\,\text{kJ/mol}$ and $\sigma = 3.768\,\text{\AA}$.

Each system is equilibrated in NPT ensemble for $100\,\mathrm{ps}$ at ambient conditions, namely $T=300\,\mathrm{K}$ and
$P=1\,\mathrm{bar}$. After equilibration the box lengths of the cubic boxes are roughly $5.65\,\mathrm{nm}$ and
$7.15\,\mathrm{nm}$ for the small (6000 SPC/E) and big (12000 SPC/E) setups respectively. In order to gather enough
statistics to probe long-time dynamics within the thin solvation shell comprised of $\mathcal{O}(10)$ to $\mathcal{O}(10^2)$ water molecules
subsequent
production runs have a length of $200\,\mathrm{ns}$ for the small systems and $100\,\mathrm{ns}$ for the big systems with a time
step of $2\,\mathrm{fs}$ and writing period of $200\,\mathrm{fs}$. Energy divergence is prevented by simulation
in NVT ensemble
($T=300\,\mathrm{K}$) applying the Nos\'e-Hoover thermostat with a coupling period of $1\,$ps. The Nos\'e-Hoover is
implemented by an
extra degree of freedom introduced as a heat bath within the hamiltionian of the simulation and thus has the advantage of
creating physically more realistic NVT ensembles \cite{Nose,Hoover}. 

Additional simulations for the cavities with $r_0 = 0\,\text{\AA}$, $1\,\text{\AA}$, $2 \,\text{\AA}$ and $4\,\text{\AA}$
were performed in order to resolve the temperature dependence of our observations. Production runs with $T = 260\,$K, $280\,$K and $320\,$K were
performed up to 100$\,$ns.

All simulations are performed with the \textsc{GROMACS} 4.5.4 package \cite{gromacs}.

\subsubsection{Planar Limit}

\begin{figure}[t]
\centerline{\includegraphics[scale=1]{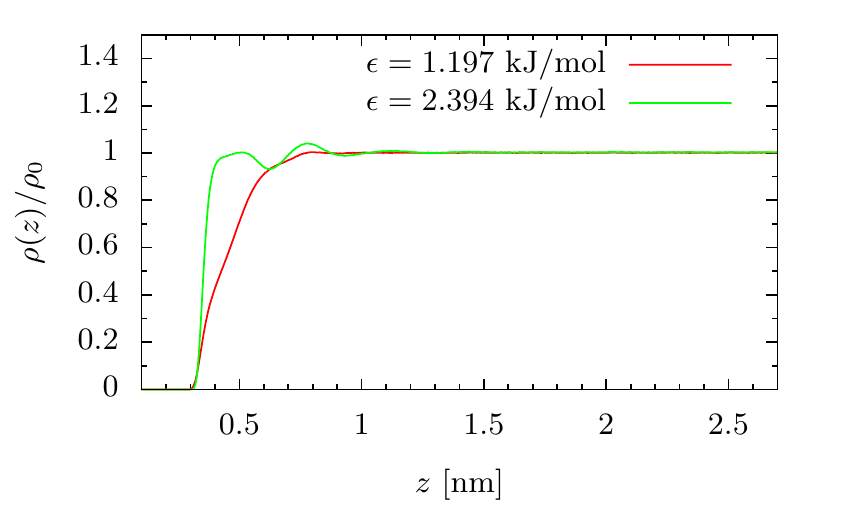}}
\caption{Normalized water distribution at hydrophobic walls.}
\label{fig1}
\end{figure}

The spatial distribution function in figure \ref{fig1} of water at planar walls with the interaction parameters mentioned above does not exhbit
a clear first and second minimum. The missing extrema hinder a clear definition of a first and second solvation shell as it can
be observed from the RDFs (main text Fig. 1) for curved surfaces.

One further simulation of a wall interacting with a doubled interaction strength $\epsilon=2.394\,\mathrm{kJ}/\mathrm{mol}$ was
conducted to generate clear extrema (Fig. \ref{fig1}) and hence definitions for the solvation layers. Clearly, the found borders
to the
solvation shells only approximate values that correspond to values that would consistently define hydration layers for the
original interaction parameters.

Another notable difference to simulations of water at curved surfaces arrises from the lack of long-range correction for the
interaction in the planar geometry. This creates
smaller water densities and larger mobilities than in bulk \cite{Sedlmeier&Netz}. 

Both discrepancies introduce minor consistency deviations between the simulations with the curved surfaces and the planar
limit but never conflict with our interpretation.

\subsection{MFPT method and perpendicular diffusivity}

In the main text we briefly describe the MFPT method from Hinczewski \textit{et al.}~\cite{Hinczewski&Netz} which takes the free
energy profile $F(r)$ and the spatial derivative of the MFPT curve $T_{fp}(r,r_t)$. Both are obtained equivalently to the
application of pair diffusion of water and methane by Sedlmeier \textit{et al.}~\cite{Sedlmeier&Netz}. The
integral $\int_{r_{min}}^{r} \mathrm{d}r\,\mathrm{e}^{-\beta F(r)}$ is evaluated numerically starting from the reflective
boundary $r_{min}$, which we define by the largest distance from the respective solute where $F(r_{min}) \geq 10$~$k_{B}$T. The
derivative $\partial\, T_{fp}(r,r_t) / \partial r$ is determined by fitting a linear function to the function values around $r-\delta r < r < r+\delta
r$ whereas $\delta r = 0.05$~nm. The closest distance $r$ evaluated for the derivative, and hence diffusivity profile value
closest to the solute, is $r \lessapprox R_{G}$ where statistics of $T_{fp}(r,r_t)$ allow a reasonable fit in the mentioned range of $\delta r$.
\begin{figure}[t]
\centerline{\includegraphics[scale=1]{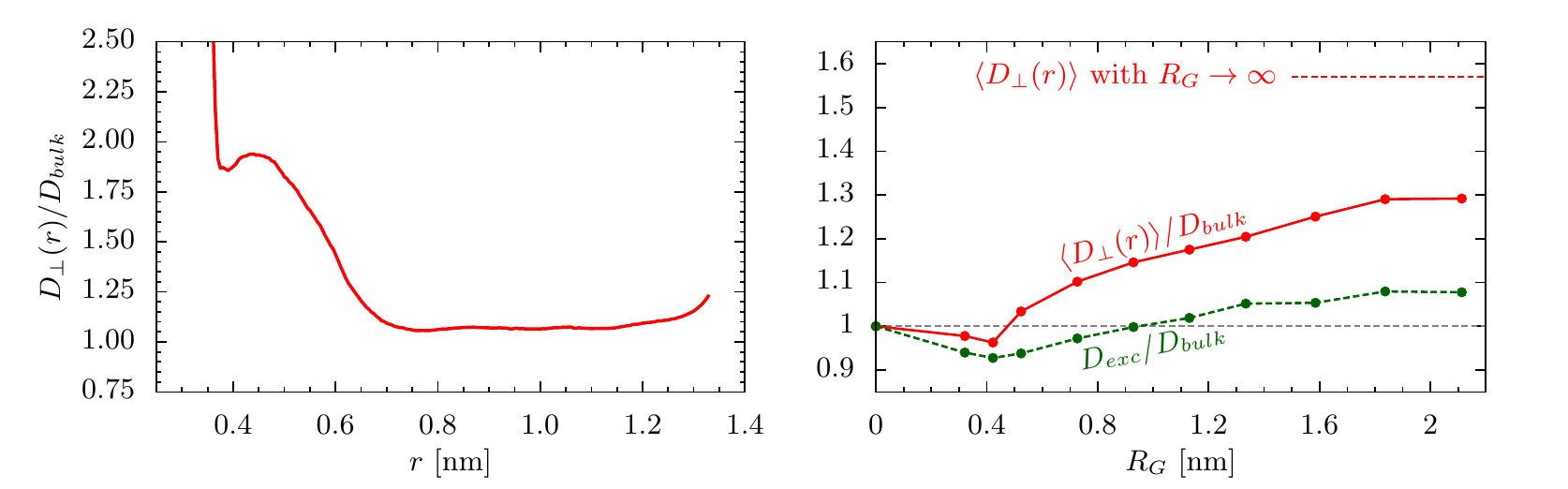}}
\caption{Left panel: Spatially resolved perpendicular diffusion profile of water at the hydrophobic wall. Right panel: (red,
solid)
Average perpendicualr diffusion constant $\langle D_{\perp}(r) \rangle$ (eq.~\ref{eq1}) inside the solvation layer of
differently sized hydrophobic model solutes. The red horizontally dashed line is the corresponding value at the planar limit
($R_G \rightarrow \infty$) for
$\langle D_{\perp}(r) \rangle$ (green, dashed) Excess perpendicular diffusion constant $D_{exc} + 1$
(eq.~\ref{eq2}) quantifying an average disruption of the diffusion of water near
hydrophobic solutes. The reference value of bulk diffusivity is drawn as grey horizontally dashed line as isoline with value 1.}
\label{fig2} 
\end{figure}

The left panel of Fig.\ref{fig2} plots the diffusivity profile of water motion perpendicular to the hydrophobic planar
model surface. The diffusivity right at the surface decreases rapidly from high mobility values to a global minimum proceeding
into a maximum which is located within the hydration water layer. With growing distance to the wall the diffusivity then
decreases to bulk diffusion. The minimum at the outer margin of the solvation shell which is visible in each profile in Fig.4 of
the main text almost fully vanishes for the profile of the planar surface here.  

The right panel in Fig.\ref{fig2} shows two further measures quantifying the perpendicular diffusivity in vicinity of the hydrophobic model solutes. The red curve is an average diffusivity inside the first solvation shell 
\begin{equation}
\langle D_{\perp}(r) \rangle = \int_{R_{max}}^{R_1} \mathrm{d}r\,g_{so}(r)\cdot D_{\perp}(r) \,\,\, .
\label{eq1}
\end{equation}
It captures the non-monotonicity from the diffusivity profiles slightly decreasing below bulk diffusion near small solutes and
subsequently increases converging towards the value for perpendicular diffusion at the planar interface.

We also calculate an "excess" perpendicular diffusivity $D_{exc}$ which quantifies an average change of the diffusivity constant
in the solute's surrounding water 
\begin{equation}
D_{exc} = \int_{R_{max}}^{\infty} \mathrm{d}r\,r^2 g_{so}(r)\cdot\left [D_{\perp}(r)/D_{bulk} - 1 \right ] \,\,\, .
\label{eq2}
\end{equation}
The plot shows $1 + D_{exc}(R_G)$. Certainly the non-monotonic curvature dependence in the diffusivity profiles is
smoothly captured in this quantity aswell. In general an excess diffusivity constant can be of particular interest in order to
estimate water diffusivity changes in dilute solutions of hydrophobic solutes.

\begin{figure}[t]
\centerline{\includegraphics[scale=1]{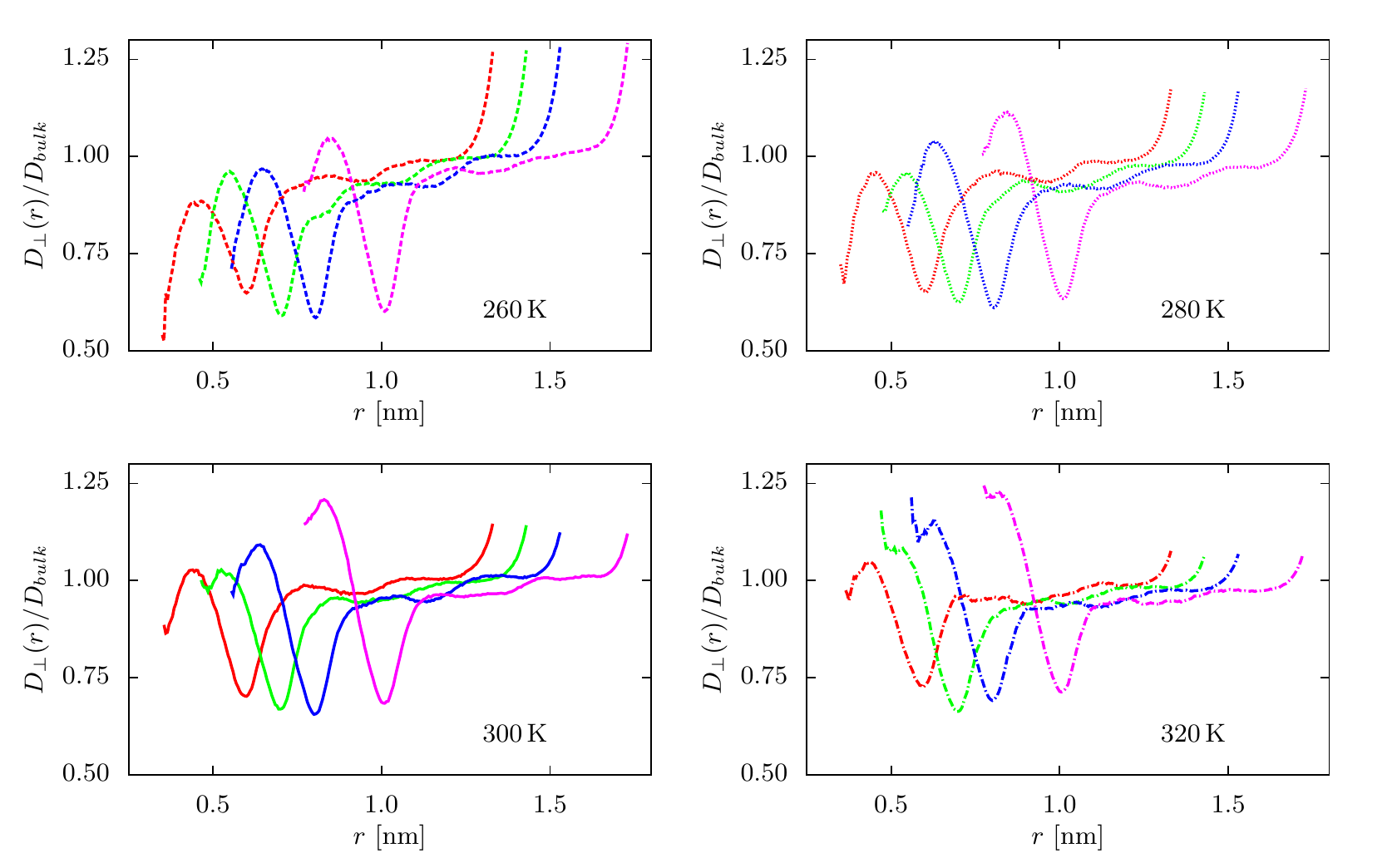}}
\caption{Perpendicular diffusion at cavities with $r_0 = 0\,\text{\AA}$, $1\,\text{\AA}$, $2\,\text{\AA}$ and $4\,\text{\AA}$ with temperatures $T=260\,$K (upper left), $T=280\,$K (upper right), $T=300\,$K (bottom left) and $T=320\,$K (bottom right).} 
\label{fig3}
\end{figure}

\subsection{Hydrogen Bond Time Correlation}

We analyse the autocorrelation function $c(t)$ of the H-bond existence inside the first hydration layer using an in-house analysis code.
It uses a geometric H-bond definition with a minimum donor-acceptor distance $d_{HB} \leq 3.5$~\AA~ and
donor-hydrogen-acceptor angle $\theta_{HB} \geq 150^{\circ}$. 

\begin{figure}[t]
\centerline{\includegraphics[scale=1]{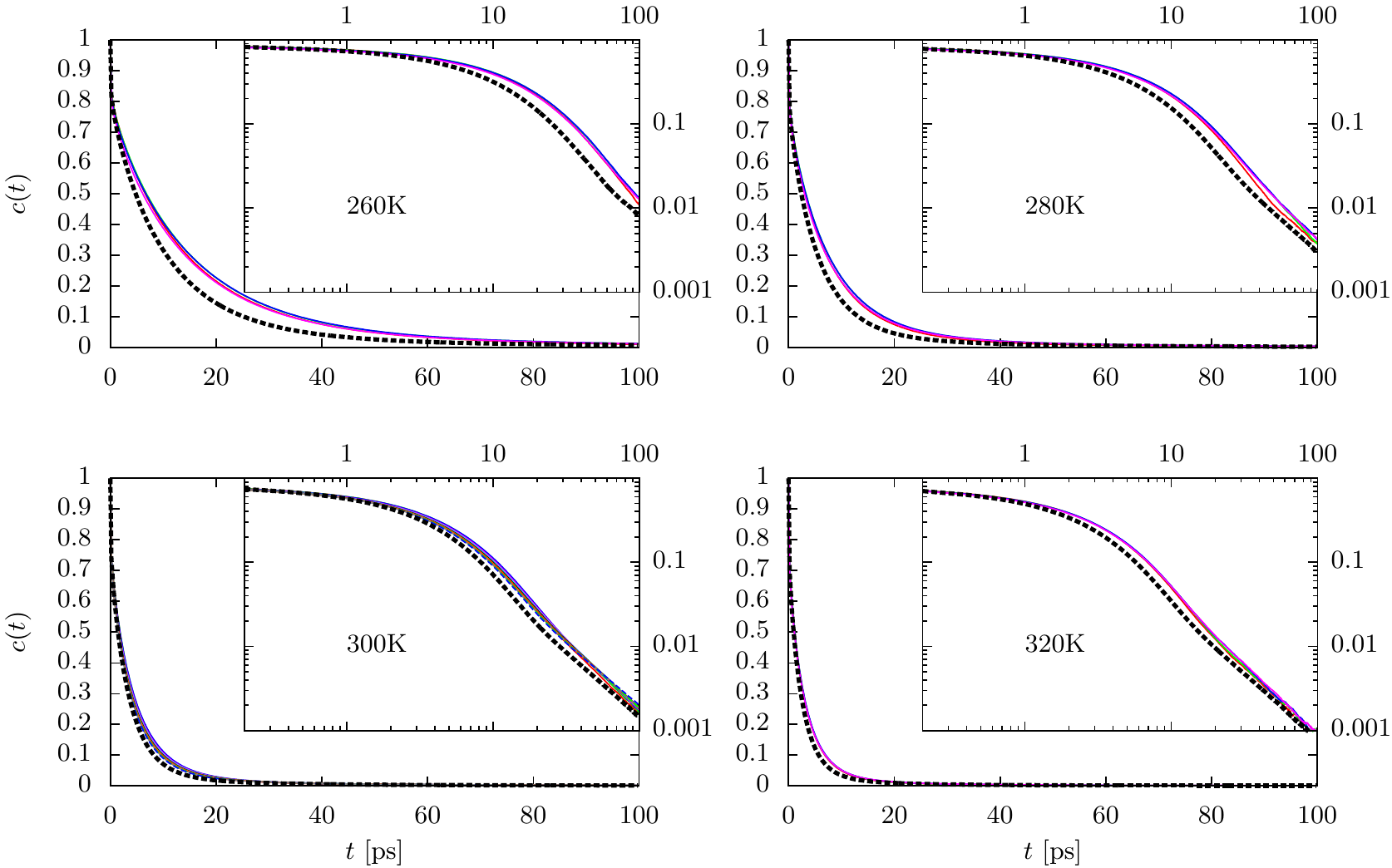}}
\caption{Correlation function $c(t)$ of the hydrogen bonding operator $h(t)$ of hydrogen bonds on water molecules inside the first solvation shell with temperatures $T=260\,$K (upper left),
$T=280\,$K (upper right), $T=300\,$K (bottom left) and $T=320\,$K (bottom right). Only at $T=300\,$K all cavity sizes were evalutated. For the
rest the analysis was limited to cavities with $r_0 = 0\,\text{\AA}$, $1\,\text{\AA}$, $2\,\text{\AA}$ and $4\,\text{\AA}$.} 
\label{fig4}
\end{figure}

Fig.~\ref{fig4} plots the correlation functions $c(t)$ for 
hydrogen bonds inside the first hydration layer.
Hydrogen bonds were selected at time 0 if they were intact and both, acceptor and donor molecules, resided in the first hydration shell.
We evaluate $c(t)$ up 
to correlation times of 100~ps until when it 
has decayed to
magnitudes of $10^{-3}$. 
Therefore, the integral over the correlation function in this time window provides a reasonable lower boundary to the actual HB
lifetimes. 

In addition the average number of hydrogen bonds per water molecule $n$ within the solvation layer was counted and presented in
the main text. The error of the average number of hydrogen bonds per molecule is estimated by block averaging. Using a frequency
of a $1000\,$ps the average number of hydrogen bonds within a block of $100\,$ps length was calculated, giving $100$
uncorrelated block averages $n_{Block}$. These were used to approximate the standard deviation error by $\delta n =
[ \Sigma_{Blocks} \,(n - n_{Block})^{2}/100 ]^{1/2} $.

\subsection{Temperature dependence of hydrogen bond dynamics}

The hydrogen bond correlation function $c(t)$ contains all three time scales of the reaction-diffusion model from Luzar and
Chandler~\cite{Luzar&Chandler} given by 
\begin{eqnarray*}
\frac{\partial}{\partial t} \rho (\vec{r}, t) &=&  D \nabla^2 \rho (\vec{r}, t) + \delta(\vec{r})\, k c(t) - \delta(\vec{r})\, k' n(t) \\
 &\equiv& D \nabla^2 \rho(\vec{r},t) + k(t) \delta{\vec{r}},
\end{eqnarray*}
where the rates $k$ and $k'$ are those for breaking and reformation of a hydrogen bond, respectively. $n(t)$ is proportional to the
diffusivity since $n(t) \propto \rho(0,t)$~\cite{Luzar} (and $\rho(r,t)$ obeying Fick's law). 

Thus, as we used the reactive flux $k(t)=-\dot{c}(t)$ to evaluate the \textit{zero frequency part}
\begin{equation*}
\tau = \int c(t)\,\mathrm{d}t = \int \frac{k(t) + k' n(t)}{k}\,\mathrm{d}t \, , 
\end{equation*}
the correlation time $\tau$ contains the factor $D k' / k$, which precludes a direct one-to-one connection to the
temperature dependence of the diffusivity. Certainly a direct relation of temperature dependence between hydrogen bond
dynamics and diffusivity is restored if  the time scales were extracted separately, which, however, is not a straightforwared task.

%\begin{thebibliography}{39}
%\expandafter\ifx\csname natexlab\endcsname\relax\def\natexlab#1{#1}\fi
%\expandafter\ifx\csname bibnamefont\endcsname\relax
%  \def\bibnamefont#1{#1}\fi
%\expandafter\ifx\csname bibfnamefont\endcsname\relax
%  \def\bibfnamefont#1{#1}\fi
%\expandafter\ifx\csname citenamefont\endcsname\relax
%  \def\citenamefont#1{#1}\fi
%\expandafter\ifx\csname url\endcsname\relax
%  \def\url#1{\texttt{#1}}\fi
%\expandafter\ifx\csname urlprefix\endcsname\relax\def\urlprefix{URL }\fi
%\providecommand{\bibinfo}[2]{#2}
%\providecommand{\eprint}[2][]{\url{#2}}